\documentclass[pre,aps,amsmath,amsfonts,amssymb,notitlepage,10pt]{revtex4-1}

\usepackage{bbm} 
\usepackage{bm} 
\usepackage{graphicx}
\usepackage{xcolor}
\usepackage{CJK}

\begin{document}
\begin{CJK*}{UTF8}{gbsn} 
\title{Quantum chaos for nonstandard symmetry classes in the Feingold-Peres model of coupled tops}
\author{Yiyun Fan (范艺芸)}\affiliation{School of Mathematical Sciences,
  University of Nottingham, Nottingham NG7 2RD, UK}
\author{Sven Gnutzmann}\affiliation{School of Mathematical Sciences,
  University of Nottingham, Nottingham NG7 2RD, UK}
\author{Yuqi Liang (梁玉其)}\affiliation{School of Mathematical Sciences,
  University of Nottingham, Nottingham NG7 2RD, UK}

\begin{abstract}
  We consider two coupled quantum tops with angular momentum vectors
  $\mathbf{L}$ and $\mathbf{M}$. The coupling Hamiltonian
  defines the Feingold-Peres model, which is a known paradigm of quantum 
  chaos. We show that this model has a nonstandard symmetry with respect to 
  the Altland-Zirnbauer tenfold symmetry classification of quantum systems,
  which extends the well-known threefold way of Wigner and Dyson
  (referred to as `standard' symmetry classes here).
  We identify the nonstandard symmetry classes BD$I_0$
  (chiral orthogonal class with no zero modes), BD$I_1$
  (chiral orthogonal class with one zero mode),
  and C$I$ (antichiral orthogonal class)
  as well as the standard symmetry class A$I$ (orthogonal class).\\
  We numerically
  analyze the specific spectral quantum signatures of chaos
  related to the nonstandard symmetries. In the microscopic
  density of states and in the distribution of the lowest positive
  energy eigenvalue we show that the Feingold-Peres model
  follows the predictions of the Gaussian ensembles of random-matrix theory
  in the appropriate symmetry class if the corresponding 
  classical dynamics is chaotic. In a crossover to mixed and near-integrable
  classical dynamics we show that these signatures disappear or strongly
  change.
\end{abstract}

\date{\today}

\maketitle
\end{CJK*}

\section{Introduction}

We consider two coupled quantum tops with 
respective angular momentum operators
$\mathbf{L}=(L_x,L_y,L_z)^T$ and 
$\mathbf{M}=(M_x,M_y,M_z)^T$ and Hamiltonian
\begin{equation}
  H= \frac{1+\lambda}{j+1/2}\left(L_z+ M_z \right)+
  \frac{4 (1-\lambda)}{(j+1/2)^2}L_xM_x
  \label{Hamiltonian}
\end{equation}
where
$\lambda\in [0,1]$ is a parameter that changes the relative 
strength of the two parts,
and $j$ is the (half-integer or integer) total angular momentum
quantum number of both tops.
More than 30 years ago Feingold and Peres introduced this as a
paradigmatic model for quantum chaos \cite{fp,fmp}. We now reconsider this
model in the light of its nonstandard symmetry properties in the tenfold
symmetry classification of quantum systems by Altland and Zirnbauer
\cite{AZ1,AZ2, Zirni}. When Feingold and Peres used this model only the 
three standard symmetry classes of Wigner and Dyson 
\cite{Wigner1,Wigner2,Dyson1,Dyson2} 
had been known for quantum systems. 
The first nonstandard symmetry classes that have
been discovered by Verbaarschot and Zahed \cite{Ver1,Ver2}
are the three chiral symmetry classes (see also \cite{gade, slevin}). 
Altland and Zirnbauer then found 
the remaining four symmetry classes and revealed a direct
connection between the mathematical
structure behind this classification of quantum systems and Cartan's tenfold
classification of symmetric spaces \cite{Zirni}. In random-matrix theory
some of the relevant ensembles 
had already been known without
awareness of their application to quantum systems (e.g. the ensembles that
go back to Wishart \cite{wishart}).
The presence of a nonstandard symmetry in the Feingold-Peres model 
defined by Eq.~\eqref{Hamiltonian}
and its implication
for universal spectral properties have thus remained undetected at the time. 
Our aim is to reintroduce the model as a paradigm to understand the 
quantum signatures of chaos related to nonstandard symmetries
and numerically analyze how these signatures change
in a crossover between chaotic and integrable
motion.\\
The common feature of the seven nonstandard symmetry classes is the
presence of a mirror symmetry in the energy eigenvalue spectrum around a 
specific energy $E_F$: if $E_F+E$
is an eigenvalue, so is $E_F-E$ and they have the same degeneracy. 
Physically this is typically realized in fermionic many-body systems
such as the quasiparticle excitations in superconductors 
and superfluids described by the Bogoliubov-de Gennes
equation or relativistic Fermions described by the Dirac equation.
In these cases $E_F$ is the Fermi energy and all
energy eigenstates with energies below $E_F$ are occupied in the ground state. 
In the following we
will set $E_F=0$ without loss of generality.
The spectral mirror symmetry alters universal spectral properties such as
level-repulsion near the symmetry energy $E=0$ and the effects die off quickly
on the scale of the mean level spacing. Far away from the symmetry point
the standard behavior of the three Wigner-Dyson symmetry classes
is recovered.
The nonstandard symmetry classes are well studied in 
random-matrix theory and physically more realistic disordered systems
such disordered superconductors \cite{AZ1,AZ2} or 
Dirac particles in a disordered gauge field \cite{Ver1,Ver2}.
The Feingold-Peres model is an example of a quantum system where the
nonstandard symmetry classes are realized in a finite Hilbert space.
In such a setting one may describe the nonstandard symmetry
classes in terms of anticommuting operators: 
the spectral mirror symmetries
may be described in terms of unitary (chirality) or antiunitary 
(charge-conjugation) operators $C$
such that $HC+CH=0$, or $CH C^{-1}=-H$. 
It is sufficient to consider $C^2=\pm 1$. 
Considering all possible (nonredundant) 
combinations of the various types (or nonexistence) 
of operators $C$ with the threefold classification with respect to
the behavior under time-reversal transformations 
leads to the tenfold classification, see \cite{stargraphI}.\\
For the Feingold-Peres model the operator $C$
may be identified as a unitary operator of the form
\begin{equation}
  C= e^{i \alpha} \mathcal{R}^{(1)}_x(\pi) \otimes \mathcal{R}^{(2)}_y(\pi) =
  e^{i \alpha-i \pi L_x-i \pi M_y}
  \label{eq:chiral}
\end{equation}
where $e^{i \alpha}$ is a phase factor 
that has no effect in the present discussion but will need to be 
included later.
This describes a rotation of the first top by an angle $\pi$ 
around the x-axis and
a rotation of the second top by the same angle about the y-axis.
As a transformation this changes the sign of all angular
momentum components that are not aligned with the axis of rotation. In
particular one has 
$C L_z C^{-1}=-L_z$, $C M_z C^{-1}=-M_z$, 
$C L_x C^{-1}=L_x$ (no sign change here), and $C M_x C^{-1}=-M_x$.  
Altogether one then has
\begin{equation}
  C H C^{-1}
  = -H
  \label{chiral_symmetry}
\end{equation}
which implies nonstandard symmetries.
As the Hamiltonian is also invariant under two unitary symmetries
(exchange of the two tops $\mathbf{M} \leftrightarrow \mathbf{L}$ and
rotations of both tops by $\pi$ around the $z$ axis)
the detailed classification needs to be done carefully, considering
each invariant subspace separately. In Section~\ref{sec:symmetry} 
we will show that different invariant subspaces realize
different symmetry classes and that the classification also depends
on whether $j$ is integer or half-integer.
We will also show there that the Hamiltonian has a generalized
time-reversal symmetry.

In random-matrix theory 
statistical correlations in the eigenvalue spectra 
for all symmetry classes have been studied 
with great detail using Gaussian ensembles. 
This is a powerful tool for disordered systems and quantum chaos
as the nontrivial  
predictions of the Gaussian ensembles are known to be universal
(with varying degree of rigor depending on the setting).
Being deliberately vague the corresponding universality
classes comprise quantum systems with `sufficient complexity'.
Universal statistics of spectral fluctuations means that
appropriately averaged spectral functions in physical systems
are independent of the details of the system and only depend on
the symmetry classification if measured on the right energy scale.
The latter turns out to be the mean level spacing. 
Characterizing the universality classes in detail 
(specifying what `sufficient complexity' means) and
finding clear criteria whether a given system should be
universal or not is usually a very hard problem if one pursues rigor.
Numerical analysis on the other hand gives abundant evidence and examples
for universal spectral statistics.
In general terms universality can be achieved either by an
appropriate amount of disorder in the system or by the presence of chaos 
in the corresponding classical dynamics, i.e. by the presence of 
quantum chaos.

In this paper we follow the quantum chaos approach to universal
spectral statistics.
This approach
is well established for the three standard Wigner-Dyson symmetry
classes. In 1983 Bohigas, Giannoni and Schmit (BGS) \cite{BGS} conjectured 
that the spectral statistics (on the scale of the mean level spacings)
of \emph{individual} quantum chaotic systems is universal with three 
universality
classes corresponding to the three
Gaussian Wigner-Dyson ensembles of random-matrix theory (GOE, GUE, and GSE).
This conjecture was first confirmed numerically
and was a major challenge in quantum chaos for more than 20 years
\cite{diagonal,actioncorrelation,SieberRichter} 
before complete microscopic derivations using semiclassical approaches
appeared: first for quantum graph models (a quantum particle that moves freely
on a network described by a metric graph) \cite{GA1,GA2}
(see also \cite{quantumgraphs} as a review for quantum chaos on graphs)
and then for `generic' quantum chaotic systems \cite{Essen}
(see \cite{SuSyreview} and references therein for a review). 
Hopes for a rigorous 
and general proof had to be abandoned due to the existence of `ungeneric'
counter examples \cite{catmaps,arithmetic}.
Still, one may hope for rigorous proofs for certain
well-defined quantum models (e.g. quantum graphs).\\
The BGS conjecture may directly be extended to 
nonstandard symmetry classes (see Section~\ref{BGS})
where the quantum chaos approach has so far been
restricted to quantum graph models \cite{stargraphI, stargraphII}
and to Andreev billiards \cite{MagneticAndreev}.
The nonstandard symmetries have a direct and universal impact on
the (microscopic) density of states near $E=0$. 
For both, quantum graphs and quantum chaotic magnetic Andreev billiards 
a periodic-orbit analysis (the self-dual approximation)
based on Gutzwiller-type semiclassical trace formulas is consistent 
with the generalized BGS conjecture. 
As is well known \cite{kosztin} nonmagnetic Andreev billiards 
cannot show proper quantum chaos (in the sense that the corresponding 
classical dynamics is fully chaotic) and they do not belong
in the universality class of the Gaussian ensembles for the appropriate
symmetry class.
They show, however, very interesting behavior 
in their own right and may be described in terms of more
restricted Gaussian ensembles (see \cite{AndreevReview} and references therein).
For quantum graphs
all ten symmetry classes may be realized and they give numerical
evidence that supports the generalized BGS conjecture \cite{stargraphII}.\\  
We will show that the Feingold-Peres model 
is a further model where some nonstandard symmetry classes are realized
and that it can be used to study the corresponding universal signatures in
the statistics of eigenvalues.
More general models
based on coupling two tops may be used as paradigmatic systems to
study universal quantum signatures of chaos related to 
all nonstandard
symmetry classes.

Let us now give an overview of the paper.
In Section~\ref{sec:model} we introduce the Feingold-Peres model 
together with its corresponding classical dynamics and 
summarize the known crossover from chaotic to integrable dynamics
using additional numerical simulations.
In 
Section~\ref{sec:symmetry}
we analyze the proper symmetry classification of this model
and establish that different subspectra with respect to the
remaining unitary symmetries may lie in different symmetry classes.
In Section~\ref{sec:numerics} we then establish numerically that the
quantum spectra 
are consistent with the predictions of the
random-matrix ensembles of the corresponding symmetry class
and thus with the generalized BGS conjecture. This is obtained without 
any average over disorder by averaging over different representations
defined by the angular momentum 
quantum number $j$ or, equivalently different values of 
the effective Planck constant $\hbar=1/(j+1/2)$.
Section~\ref{sec:conclusions} summarizes the main conclusions
and gives an outlook on future directions.

\section{The Feingold-Peres model and its corresponding classical dynamics}
\label{sec:model}

Let us quickly summarize the theoretical setting for 
the Hamiltonian \eqref{Hamiltonian} of the Feingold-Peres model.
The fundamental quantum observables are described by the components
of angular momentum operators. They satisfy the standard commutation relations
\begin{equation}
    \left[L_k, L_{l}\right]= i \sum_{m\in \{x,y,z\}}
    \epsilon_{klm} L_m,\quad
    \left[M_k, M_{l}\right]= i \sum_{m\in \{x,y,z\}}
    \epsilon_{klm} M_m,\quad \text{and} \quad
    \left[L_k, M_{l}\right]= 0
  \label{quantum_commutator}%
\end{equation}%
where $k,l \in \{x,y,z\}$ and $\epsilon_{klm}$ is the totally antisymmetric
Levi-Civita tensor with $\epsilon_{xyz}=1$. They generate the group 
$SU(2) \times SU(2)$ and we assume that the Hilbert space $V$
is an irreducible
representation of this group such that both 
angular momenta have the same total angular
momentum $\mathbf{L}^2 =\mathbf{M}^2=j(j+1)$ and
the dimension is $N=\mathrm{dim}\ V= (2j+1)^2$. 
The standard basis will be denoted by $ \{ |m_1,m_2\rangle \}$ 
with $m_1,m_2 \in \{-j,-j+1, \dots,j\}\equiv Z_j$.
It consists of common eigenstates to
the $z$-components of both angular momenta
$L_z |m_1,m_2\rangle = m_1 |m_1,m_2\rangle$
 and $M_z |m_1,m_2\rangle = m_2 |m_1,m_2 \rangle$ .

\subsection{Classical dynamics of two coupled tops}

The classical limit of the Feingold-Peres model is well understood.
In their original work Feingold and Peres \cite{fp} observed a
crossover from chaotic to regular dynamics at energy $E=0$ 
as the system parameter $\lambda$ increases from
zero to one. In this section
we summarize the main results and add an up-to-date
numerical analysis of the crossover for comparison with
our analysis of the quantum signatures of this crossover at the
relevant
system parameters.

The corresponding classical dynamics of the Feingold-Peres model
is obtained by replacing
the rescaled quantum angular momentum operators 
$L_k/(j+1/2)$ and $M_k/(j+1/2)$ ($k=x,y,z$)
with commutator relations \eqref{quantum_commutator}
by classical angular momentum variables $\mathcal{L}_k$  and
$\mathcal{M}_k$
such that $\mathcal{L}_x^2+\mathcal{L}_y^2+\mathcal{L}_z^2=
\mathcal{M}_x^2+ \mathcal{M}_y^2+\mathcal{M}_z^2=1$.
The classical angular momentum variables thus span 
a phase space $S^2\times S^2$, the Cartesian product of two spheres.
On this phase space they obey the standard angular momentum
Poisson bracket relations
\begin{equation}
  \left\{
    \mathcal{L}_k, 
    \mathcal{L}_l
  \right\}=
  - 
  \sum_{m\in\{x,y,z\}}
  \epsilon_{klm} 
  \mathcal{L}_m ,
  \quad
  \left\{
    \mathcal{M}_k, 
    \mathcal{M}_l
  \right\}=
  - 
  \sum_{m\in \{x,y,z \} }
  \epsilon_{klm} 
  \mathcal{M}_m ,
  \quad\text{and}\quad
  \left\{
    \mathcal{L}_k, 
    \mathcal{M}_l
  \right\}= 0\ .
  \label{classical_brackets}
\end{equation}
The classical dynamics is most easily obtained by 
performing the corresponding substitutions in the 
quantum Heisenberg equations which leads Hamiltonian
equations according to
\begin{equation}
  \frac{d}{dt} L_k = i (j+1/2)\left[ H, L_k\right]
  \quad \mapsto
  \quad
  \frac{d}{dt}
  \mathcal{L}_k= \left\{ \mathcal{H}, \mathcal{L}_k\right\}
\end{equation}
and, analogously for $M_k \to \mathcal{M}_k$.
The Hamilton function for the Feingold-Peres model is  
\begin{equation}
  \mathcal{H}= \left(1+\lambda\right)\left(\mathcal{L}_z
    + \mathcal{M}_z \right)+
  4 (1-\lambda) \mathcal{L}_x \mathcal{M}_x\ .
\end{equation}
We have used units such that $\hbar=1/(j+1/2)$ which links the classical
limit $\hbar \to 0$ to the limit $j \to \infty$. It also sends the Hilbert
space dimension to infinity $N \to \infty$. The classical limit can be
constructed explicitly if one assumes that the state of the system is in
a generalized $SU(2)\times SU(2)$ coherent state and takes the expectation
value
of both sides of the quantum Heisenberg equations. 
Identifying $\mathcal{L}_k=
\frac{1}{(j+1/2)} \langle L_k\rangle$,
$\mathcal{M}_k=
\frac{1}{(j+1/2)} \langle M_k\rangle$
and taking the limit $j\to \infty$ 
then results in the classical equations of motion. Needless to say the
appearance of factors $j+1/2$ accompanying each angular momentum operator is
essential
for this limit to work.
Note however, that replacing $j+1/2$ by just $j$ (and thus $\hbar=1/j$) 
leads to the same limit -- we write $j+1/2$ merely because this is the more
natural expansion in the semiclassical regime.\\
As the product of two spheres the classical phase space has dimension four. 
There are many ways to introduce two pairs of canonical coordinates
$(P_1,Q_1,P_2,Q_2)$ 
with canonical Poisson brackets $\{P_\ell,Q_{\ell'} \}=\delta_{\ell \ell'}$ and 
$\{P_{\ell},P_{\ell'} \}=\{Q_{\ell},Q_{\ell'} \}=0$
on this phase space. While any choice necessarily
leads to some coordinate singularity we found the choice 
\begin{subequations}
  \begin{align}
    \mathcal{L}_x=&\sqrt{1-P_1^2}\cos Q_1,&  
    \mathcal{M}_x=&\sqrt{1-P_2^2}\cos Q_2,\\
    \mathcal{L}_y=&\sqrt{1-P_1^2}\sin Q_1,   &
    \mathcal{M}_y=&\sqrt{1-P_2^2}\sin Q_2, 
    \\
    \mathcal{L}_z=&P_1,&
    \mathcal{M}_z=&P_2
  \end{align}
\end{subequations}
most convenient.

\begin{figure}
  \includegraphics[width=0.4\textwidth]{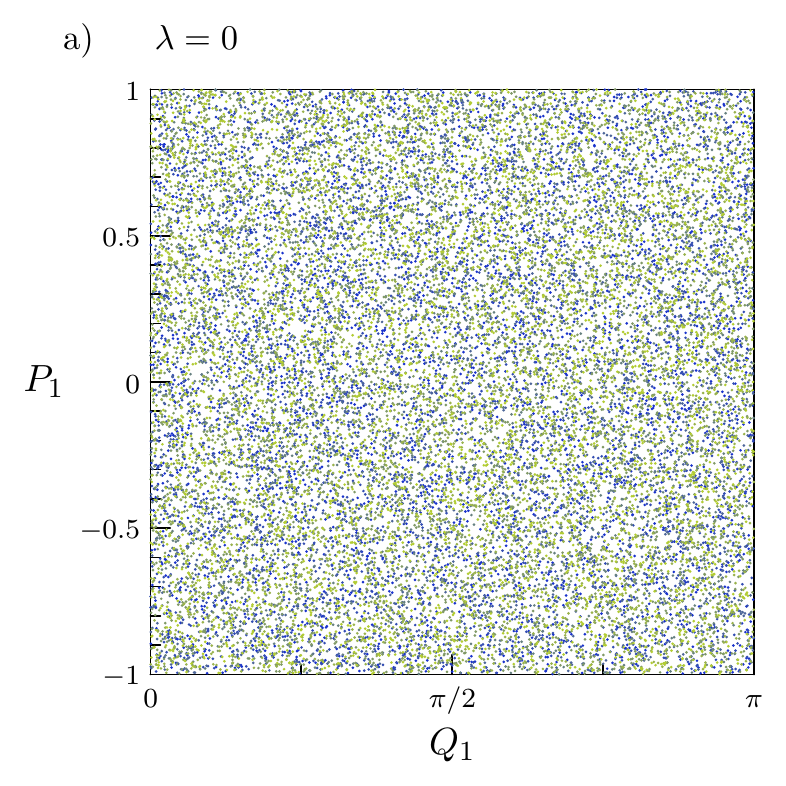}
  \includegraphics[width=0.4\textwidth]{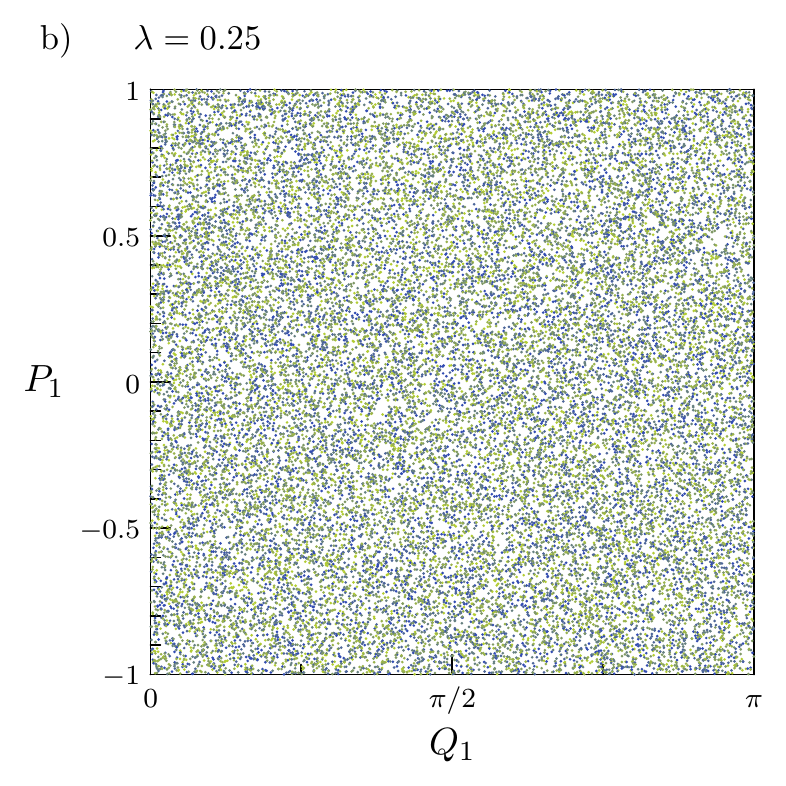}\\
  \includegraphics[width=0.4\textwidth]{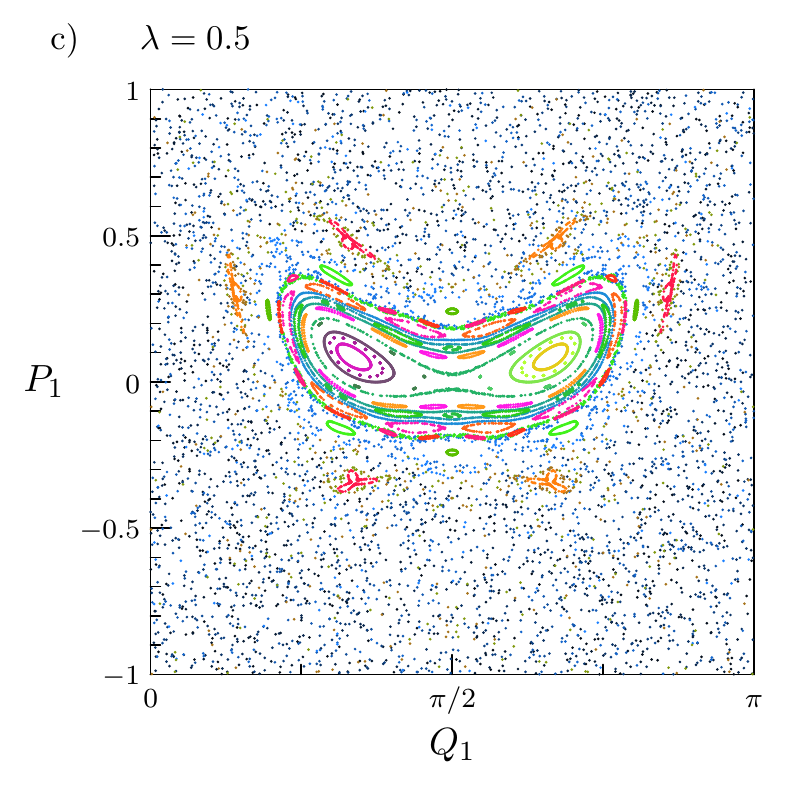}
  \includegraphics[width=0.4\textwidth]{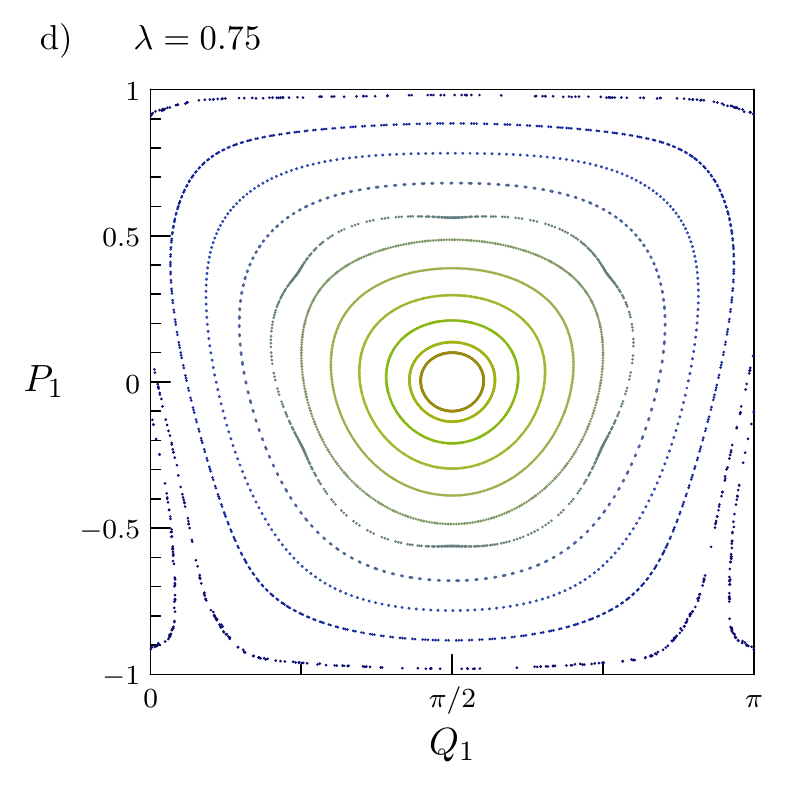}
  \caption{(Color online) Poincar\'e plots in the energy shell $E=0$
    for various values of the coupling parameter $\lambda$.
    For $\lambda=0$  (graph a)) and $\lambda=0.25$ (graph b)) 
    the Poincar\'e plots show no
    structures indicating that the dynamics is dominantly chaotic. While
    very small regular islands cannot be excluded they would have to be
    on scales much smaller than the resolution of this plot. At $\lambda =0.5$
    (graph c)) the plot shows a typical mixed phase space with a large 
chaotic region
    and a regular island. At $\lambda=0.75$ (graph d)) the plot shows a near 
    integrable dynamics.
    \label{fig1}}
\end{figure}

For $\lambda=1$ the classical dynamics is clearly
integrable with two constants of motion $\mathcal{L}_z$ and $\mathcal{M}_z$.
Peres and Feingold have shown that, as $\lambda$ is decreased, 
there is a crossover to mixed and chaotic
dynamics on the energy shell $\mathcal{H}\equiv E=0$ 
for intermediate values and chaos being maximally developed at
$\lambda =0$. In the remainder of this section we summarize 
the relevant findings using our own numerical calculations.\\
The transition can most easily be seen with Poincar\'e plots as 
shown in Fig.~\ref{fig1}. 
These were obtained by numerically integrating the classical equations of 
motion with initial conditions that satisfy $\mathcal{H}(P_1,Q_1,P_2,Q_2)=0$
and $Q_2=\frac{\pi}{2}$. The latter condition defines our
choice for the Poincar\'e surface of section. 
Whenever the trajectory crosses the
Poincar\'e surface we record the coordinates $Q_1\in[0,2\pi)$ and 
$P_1\in[-1,1]$ that describe
the orientation of the first top. The symmetries of the dynamics allow
us to reduce this Poincar\'e map to $Q_1\in[0,\pi)$ and 
$P_1\in[-1,1]$ which is shown in Fig.~\ref{fig1}. The 
remaining mirror symmetry about $Q_1=\pi/2$ of Poincar\'e plots
is related to the time-reversal symmetry of the dynamics. \\
We do not show any plots for $0.75<\lambda < 1$ as the graph only 
changes slightly. In the limit $\lambda \to 1$ the lines follow
the level curves of $\mathcal{L}_y=\sqrt{1-P_1^2}\sin Q_1$ and this is
already visible clearly at $\lambda=0.75$. At first sight one may expect
to see the level lines of $P_1$ as the Hamiltonian
at $\lambda=1$ reduces to
$\mathcal{H}=2(P_1+P_2)$ which has obvious constants of motion $P_1$ and
$P_2$. While this is true the dynamics is indeed highly degenerate
as both tops rotate 
at the same frequency with constant
$z$-components of their angular momentum vectors. This implies that the
Poincar\'e map maps each point $(Q_1,P_1)$ back to itself and that each 
trajectory in the energy shell $E=0$ is a periodic orbit with same period.
In such a situation the perturbation that sets in for $\lambda < 1$ 
defines the invariant tori visible in a Poincar\'e map and a 
simple perturbative calculation shows that the Poincar\'e map follows 
level lines of $\mathcal{L}_y$.

While the Poincar\'e plots give a clear qualitative picture of the
crossover from chaotic to integrable motion let us also give a more
quantitative analysis of the degree of chaos by considering the exponential
dependency on initial conditions as measured by the Lyapunov exponents.
For this we numerically calculate 
\begin{equation}
  \Lambda_\epsilon(t)= \frac{1}{t} \log \frac{\Delta_\epsilon(t)}{\epsilon} 
\end{equation}
where $\Delta_\epsilon (t)= \sqrt{\Delta {P_1}^2 + \Delta {Q_1}^2 + \Delta
  {P_2}^2+ \Delta {Q_2}^2  }$ is the distance between two trajectories
with initial distance $\Delta_\epsilon (0)=\epsilon>0$. 
The Lyapunov exponent is obtained in the limit
$\Lambda = \lim_{t\to \infty} \lim_{\epsilon \to 0} \Lambda_\epsilon(t)$
(where the order of the two limits is important). 
Numerically one has to work with a finite but small
value for $\epsilon$ (in our case $\epsilon \approx 10^{-12}$).
For any finite value of $\epsilon$ one always has $\Lambda_\epsilon(t)\to 0$
as $t \to \infty$ and to estimate the Lyapunov exponent one has
to use a cutoff time $t_c$ 
that ensures that the deviation $\Delta_\epsilon (t_c)$ is still small
compared to the extension of phase space.
\begin{figure}
  \includegraphics[width=0.4\textwidth]{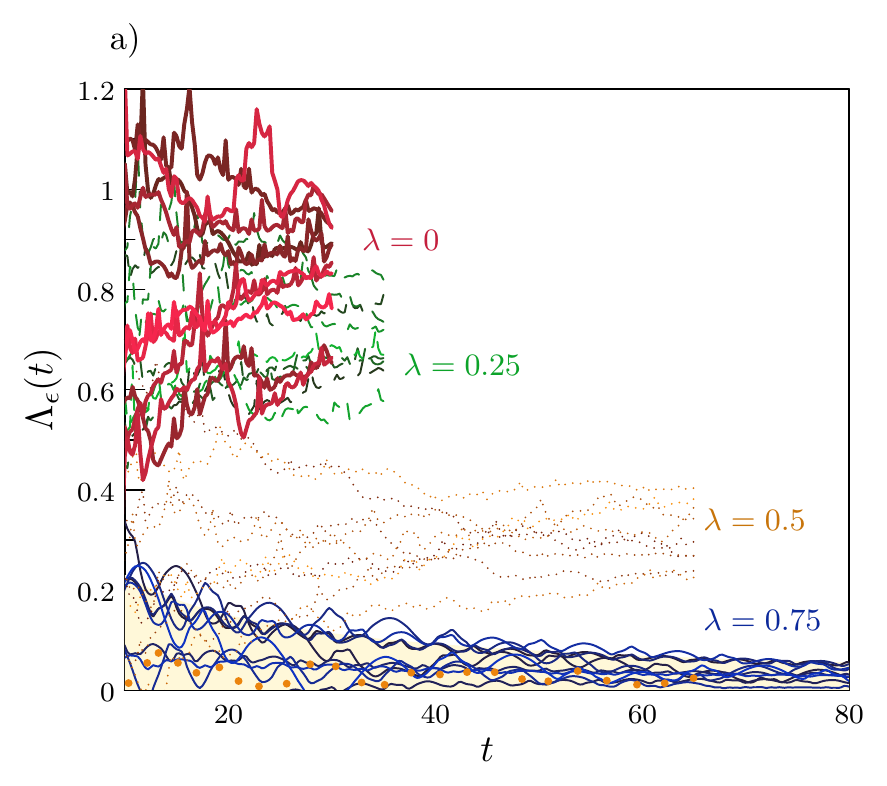}
  \includegraphics[width=0.4\textwidth]{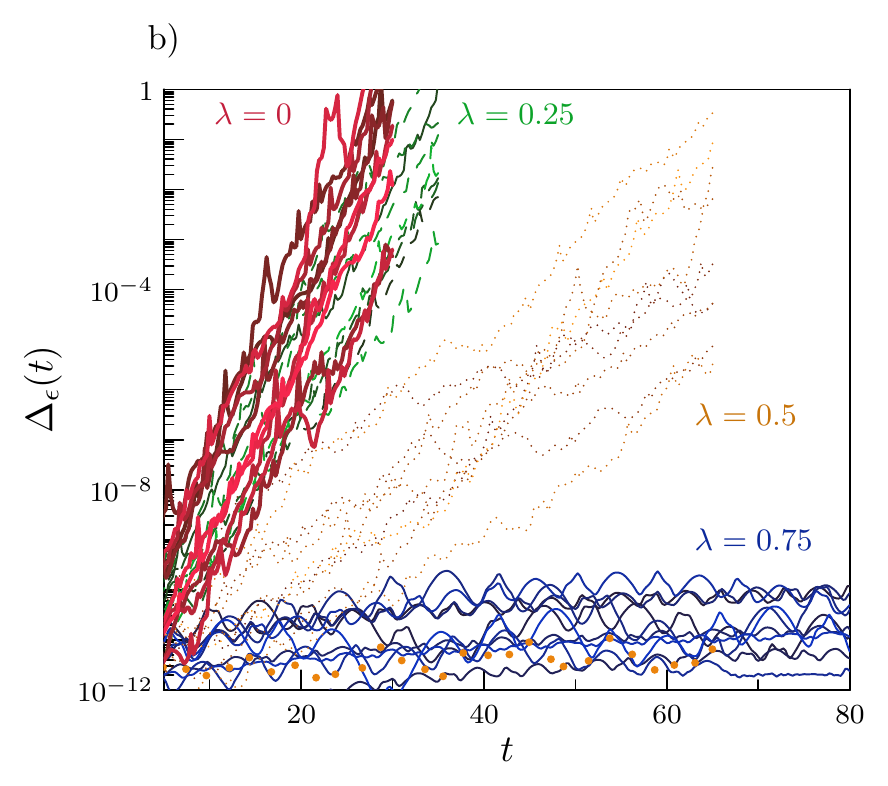}
  \caption{(Color online) Exponential dependence on initial conditions as measured by
    $\Lambda_\epsilon(t)$ (left graph a)) and $\Delta_\epsilon (t)$ 
    (logarithmic plot in right graph b)). 
    These are shown for ten trajectories for each of the values
    $\lambda = 0$ (fat red lines), $\lambda=0.25$ (dashed green lines),
    $\lambda=0.5$ (dotted yellow lines), and $\lambda=0.75$ (thin blue lines).
    For $\lambda=0.5$ 9 trajectories are in the chaotic region 
    (thin dotted lines) and one is inside the regular island (fat dotted
    line).
    Cutoff times were chosen such that most trajectories for the same value
    $\lambda$ obey $\Delta_\epsilon(T_c) \ll 1$.
    \\
    The shaded region on the left graph for 
    $\Lambda_\epsilon (t)$ is bounded by $\frac{\log
      t}{t}$
    which gives the typical decay of $\Lambda_\epsilon(t)$
    when the Lyapunov exponent vanishes.
    \label{fig2}}
\end{figure}
In Fig.~\ref{fig2}
we show plots for $\Lambda_\epsilon(t)$ 
and $\Delta_\epsilon (t)$ for a number of trajectories for different
values of the coupling parameter $\lambda$.
The findings are consistent with the Poincar\'e plots in Fig.~\ref{fig1}.
For $\lambda=0$ we find chaotic dynamics as can clearly be seen
in the exponential increase of $\Delta_\epsilon (t)$.
Though the Lyapunov exponents have not yet converged one may give an
estimate $0.65 < \Lambda < 1$.
For $\lambda=0.25$ the dynamics is also chaotic but the exponential 
increase is a bit weaker $0.55< \Lambda< 0.85$. For $\lambda=0.5$
the mixed phase space shows up in that some trajectories
show exponential dependency on initial conditions
with $0.2 < \Lambda < 0.4$ while others
do not show this and the numerics is consistent with $\Lambda=0$.
For the ten plotted trajectories initial conditions were
chosen random and nine turned out to be chaotic and one
regular. We have checked consistency with the 
corresponding Poincar\'e plot.\\
It is certainly possible to improve these numerical results with
some effort. For our purpose, which is to investigate 
quantum signatures of chaos in the spectral fluctuations
close to $E=0$ we have given a sufficient summary of
the corresponding classical dynamics.

\section{Reduction to invariant subspaces and symmetry classification} 
\label{sec:symmetry}

Let us now come back to the quantum dynamics and reduce the system
using all available unitary symmetries before considering the
symmetry classification 
of the Hamiltonian in the reduced subspaces of 
the Hilbert space $V$ according to the Altland-Zirnbauer tenfold way. 

\subsection{Unitary symmetries and their invariant subspaces}

The Hamiltonian \eqref{Hamiltonian} is invariant under exchange of the
two tops and under a simultaneous rotation of both tops by 
an angle $\pi$ around the $z$-axis. 
Let us denote the corresponding unitary quantum operators by 
$U_1$ and $U_2$ and define them through their action in the standard basis
\begin{subequations}
  \begin{align}
    U_1 | m_1, m_2\rangle =&|m_2,m_1\rangle\\
    U_2 |m_1,m_2\rangle =& e^{i\pi(2j -m_1 -m_2)}|m_1,m_2\rangle
  \end{align}
\end{subequations}
for $m_1,m_2\in Z_j$. So $U_1$ 
describes the exchange of tops. It has eigenvectors
$|m_1,m_2\rangle \pm |m_2,m_1\rangle$ with eigenvalues $\pm 1$
and thus obeys $U_1^2=1$. It acts on angular momentum operators as
$U_1 L_k U_1^\dagger= M_k$ and
$U_1 M_k U_1^\dagger= L_k$.
The operator $U_2$ is already diagonal in the standard basis.
Note that we have chosen $U_2=e^{i 2j \pi}\mathcal{R}_z(\pi)
= \pm \mathcal{R}_z(\pi)$ 
where $\mathcal{R}_z(\pi)=\mathcal{R}_z^{(1)}(\pi) \otimes \mathcal{R}_z^{(2)}(\pi)=  
e^{-i\pi \left(L_z+M_z\right)}$ describes the simultaneous rotation of both 
tops. The additional scalar factor does not change the action on 
angular momentum
operators that leaves the $z$ components invariant and inverts
the $x$ and $y$ components.
The additional scalar factor does however 
affect the eigenvalues which
are $e^{i\pi(2j -m_1 -m_2)}= \pm 1$. This implies $U_2^2=1$ as well.
The two operations clearly commute $U_1U_2= U_2 U_1$
with the common eigenbasis states $\propto  \left( 
|m_1,m_2\rangle \pm |m_2,m_1\rangle\right)$.
Their action on angular momentum operators implies that they both leave the 
Hamiltonian invariant
\begin{equation}
  U_1 H U_1^\dagger= U_2 H U_2^\dagger=H \ .
\end{equation}
The Hilbert space thus splits into four invariant orthogonal 
$V=V_{++}\oplus V_{+-} \oplus V_{-+} \oplus V_{--}$
and the Hamiltonian becomes block-diagonal in the 
appropriately ordered common eigenbasis
\begin{equation}
  H= \begin{pmatrix}
    H_{++}& 0 & 0 & 0\\
    0 & H_{+-} & 0 & 0\\
    0 & 0 & H_{-+}& 0\\
    0 & 0 & 0 & H_{--}
  \end{pmatrix}\ .
  \label{Blockhamiltonian}
\end{equation}
For later use we describe the subspaces in a little more 
detail. The eigenbases of the invariant subspaces are given by
\begin{subequations}
  \begin{align}
    |m_1,m_2,++\rangle =
    &
      \begin{cases}
        |m_1,m_1\rangle & \text{if $m_1=m_2$,}\\
        \frac{|m_1,m_2\rangle + |m_2,m_1\rangle}{\sqrt{2}}
        & \text{if $m_1<m_2$,}
      \end{cases}\\
    & 
      \text{where $m_1\in Z_j$, $m_2\ge m_1$ 
      and $2j-m_1-m_2$ is even;}\nonumber \\[0.2cm]   
    |m_1,m_2,+-\rangle =
    &
      \frac{|m_1,m_2\rangle + |m_2,m_1\rangle}{\sqrt{2}}\\
    & 
      \text{where $m_1\in Z_j$, $m_2> m_1$ 
      and $2j-m_1-m_2$ is odd;}\nonumber \\[0.2cm]   
    |m_1,m_2,-+\rangle =
    &
      \frac{|m_1,m_2\rangle - |m_2,m_1\rangle}{\sqrt{2}}\\
    & 
      \text{where $m_1\in Z_j$, $m_2> m_1$ 
      and $2j-m_1-m_2$ is even;}\nonumber \\[0.2cm]   
    |m_1,m_2,--\rangle =
    &
      \frac{|m_1,m_2\rangle - |m_2,m_1\rangle}{\sqrt{2}}\\
    & 
      \text{where $m_1\in Z_j$, $m_2> m_1$ 
      and $2j-m_1-m_2$ is odd.}\nonumber 
  \end{align}
\end{subequations}
The corresponding dimensions are
\begin{subequations}
  \begin{align}
    N_{++}=
    &
    \begin{cases}  
      (j+1)^2  & \text{if $j=1,2,\dots$ is integer,}   \\    
      (j+1)^2-\frac{1}{4} & \text{if $j=1/2,3/2,\dots$ is half-integer,}
    \end{cases}
    \\
    N_{+-}=&(2j+1)(j+1)-N_{++},\\
    N_{-+}=&N_{++}-2j-1,\\
    N_{--}=&N_{+-}
  \end{align}
\end{subequations}
and obey $N=(2j+1)^2=N_{++}+N_{+-}+N_{-+}+N_{--}$ as required.
Before moving on let us also state the obvious but relevant fact that
the standard basis 
$\{ |m_1,m_2\rangle\}$ and the common eigenbasis
$\{|m_1,m_2,\pm,\pm\rangle\}$ are related by a real orthogonal transformation.

\subsection{Symmetry classification of the reduced Hamiltonians}

Once all unitary symmetries have been taken into account to reduce a 
quantum mechanical system there remains a tenfold symmetry classification 
of the reduced system in terms of time-reversal invariance and spectral 
mirrors symmetries.
Let us start with considering time-reversal invariance.
In the standard basis where $L_z$ and $M_z$ are diagonal 
it is well-know that the matrices for $L_x$, $M_x$, $L_z$
and $M_z$
are real symmetric while the matrices of $L_y$ and $M_y$ 
are purely imaginary and 
antisymmetric. Let us thus define the antiunitary operator $T$ by its action
on an arbitrary state expanded in the standard basis
\begin{equation}
  T \sum_{m_1,m_2 \in Z_j} \alpha_{m_1,m_2} |m_1,m_2\rangle=
  \sum_{m_1,m_2 \in Z_j} \alpha_{m_1,m_2}^* |m_1,m_2\rangle\ .
\end{equation}
We see that $T$ is equivalent to the complex conjugation operator 
with respect to this basis and it squares to the identity $T^2=1$
such that $T^{-1}=T$.
Moreover, by considering matrix elements it is straight forward to see
\begin{subequations}
  \begin{align}
    TL_xT^{-1}=&L_x,&
    TL_yT^{-1}=&-L_y,& TL_zT^{-1}=&L_z,\\
    TM_xT^{-1}=&M_x,&
    TM_yT^{-1}=&-M_y,& TM_zT^{-1}=&M_z .
  \end{align}
\end{subequations}
The operator $T$ is thus an nonconventional time-reversal operator
(the conventional time reversal operator changes the sign of all 
angular momentum 
components) and the Hamiltonian is invariant under time-reversal
\begin{equation}
  [T,H]=0 \ .
\end{equation}
This time-reversal invariance carries over to all invariant subsystems
because a real orthogonal matrix has been used to obtain the block-diagonal 
form of $H$. We will use the same symbol $T$ for the induced operation
in the subspaces
(complex conjugation with respect to the common basis 
$\{|m_1,m_2,\pm,\pm\rangle\}$) and thus have
\begin{equation}
  [H_{++},T]=0, \quad [H_{+-},T]=0, \quad
  [H_{-+},T]=0, \quad \text{and} \quad 
  [H_{--},T]=0\ .
\end{equation}
In the tenfold symmetry classification there are three symmetry classes
with invariance under a time-reversal operation that obeys
$T^2=1$:
A$I$ also known as the orthogonal symmetry class;
BD$I$ also known as the chiral orthogonal symmetry class; and
C$I$ which we will call the antichiral orthogonal class
for reasons to become clear shortly (this is not an established name).\\
The difference between these three classes is the existence of
a unitary operator $C$, that we will call chirality operator,
such that Eq.~\eqref{chiral_symmetry}, or equivalently
$CH+HC=0$, holds
for all $H$ in the symmetry class.
In addition one requires that $C$ commutes with the time-reversal operator, 
$[C,T]=0$. 
In class A$I$ no such chirality operator can be defined.
In class BD$I$ a chirality operator exists and obeys $C^2=1$.
In class C$I$ a chirality operator exists as well but obeys 
$C^2=-1$ (hence the name antichiral class).

Let us conclude this subsection by showing that all three classes
are realized among the reduced subsystems in the present
model of coupled tops. 
In \eqref{eq:chiral} we have defined a chirality
operator $C$ with an unspecified phase $\alpha$.
Setting $\alpha= \pi j$ this becomes
\begin{equation}
  C=e^{i \pi j} 
  \mathcal{R}^{(1)}_x(\pi) \otimes \mathcal{R}^{(2)}_y(\pi) =
  e^{i\pi \left(j - L_x-M_y\right)} .
\end{equation}
Note that $C$ acts on the full Hilbert space $V$. 
The scalar factor $e^{i \pi j}$ 
ensures $[C,T]=0$. We also find
\begin{equation}
  C^2=\begin{cases}
    1 & \text{if $j=1,2,\dots$ is integer}\\
    -1 & \text{if $j=1/2,3/2,\dots$ is half-integer.}
  \end{cases}
\end{equation}
This seems to indicate that the system is either in the chiral  
orthogonal or antichiral orthogonal class depending on the 
angular momentum quantum 
number $j$. However we also need to check that $C$ induces 
appropriate chiral symmetry operators in the reduced subspaces.
Using standard properties of Wigner $D$-matrices one can show that
$C$ acts in the standard basis via
\begin{equation}
  C|m_1,m_2\rangle= (-1)^{j-m_2} |-m_1,-m_2\rangle \ .
  \label{chiral_action}
\end{equation}
In the common eigenbasis $\{ |m_1,m_2, \pm \pm\rangle\}$ where $H$ is
block-diagonal it is then straight forward to see that the 
matrix of the chirality operator 
assumes the form
\begin{equation}
  C=
  \begin{pmatrix}
    C_{++} & 0 & 0 & 0\\
    0&0&0&C_1\\
    0&0&C_{-+}&0\\
    0&C_2&0&0
  \end{pmatrix}.
\end{equation}
This shows that the global chirality operator $C$ only induces
a chirality operators inside the subspaces
$V_{++}$ and $V_{-+}$ where we have
$H_{\pm+} C_{\pm+}+C_{\pm+}H_{\pm+}=0$,
$[H_{\pm+},T]=0$, $[C_{\pm +},T]=0$.\\
Let us first assume that $j$ is integer.
We then have $C_{\pm+}^2=1$ and $H_{++}$ and $H_{-+}$
both belong to the chiral orthogonal class BD$I$. 
As $C_{\pm+}^2=1$ the eigenvalues of $C_{\pm+}$ can only be $\pm 1$.
Using \eqref{chiral_action} it is straight forward to see that for
odd $j$ the dimension $N_{++}=(j+1)^2$ of $V_{++}$ is even. 
We show in the Appendix  that 
the eigenvalues $\pm 1$ appear with the same degeneracy $N_{++}/2$.
For even $j$ however the dimension $N_{++}$ is odd and the
degeneracies cannot match exactly. 
In the Appendix we show that the difference in the 
degeneracies is always 
one in this case. The absolute value of this difference is known
as topological index $\nu$ and it 
counts the number of zero modes, which is the number of vanishing 
energy eigenvalues that may be predicted based on the anticommutation
relation alone.
The number of zero modes has a strong effect on 
chiral signatures
in chaotic or disordered spectra. The chiral orthogonal
class is therefore divided into subclasses BD$I_\nu$ with $\nu=0,1,2,\dots$
and each of the subclasses has different universal
signatures near $E=0$. 
To summarize the topological index satisfies
$\nu_{++}=0$ if $j$ is even and $\nu_{++}=1$ if $j$ is odd.
In the invariant subspace $V_{-+}$
the situation is just the other way around: whenever $j$ is odd
$N_{-+}$ is odd and $\nu_{-+}=1$ while for even $j$ one 
has $\nu_{-+}=0$.
\\
Now let us assume that $j$ is half-integer. 
Then $C_{\pm+}^2=-1$ and $H_{\pm +}$ both belong to the antichiral orthogonal
class C$I$. 
The eigenvalues of $C_{\pm +}$ are now $\pm i$. The dimensions
$N_{\pm +}$ are always even for half-integer $j$ and one can check, using
\eqref{chiral_action}, that the eigenvalues $\pm i$ have the same degeneracy
$N_{\pm +}/2$.\\
Note that
the global chirality operator does not
define any operator inside the 
subspaces $V_{+-}$ and $V_{--}$ as it maps states from one 
subspace to states in the other. For any $j$ the corresponding Hamiltonians 
$H_{+-}$ and $H_{--}$ therefore belong to the orthogonal class A$I$.

\section{Quantum chaos and numerical analysis 
  of quantum spectra}
\label{sec:numerics}

It is well known that disordered chiral (or antichiral) quantum systems 
show specific universal signatures of chirality in the vicinity 
of the energy $E=0$. Most prominently the 
averaged microscopic density of states 
shows characteristic signatures. 
To define the latter in the context of a numerical experiment
let us start with the energy spectrum 
$\{E_n\}$
that is obtained by appropriate diagonalization
of a symmetry reduced Hamiltonian.  
To analyze the specific signatures of spectral statistics
near $E=0$ it is generally not necessary to obtain a complete spectrum.
In our case we reduce the Hamiltonian \eqref{Hamiltonian}
into its four reduced blocks $H_{\pm \pm}$ as given by 
\eqref{Blockhamiltonian}.
For each block
we numerically evaluated $M=60$ (for even dimensional invariant subspaces) 
or $M=61$ eigenvalues (for odd dimensional invariant subspaces)
$\{E_n\}$ close to $E=0$ 
for a large number of
representations defined by the quantum number $j$ (all integer and
half-integer numbers from $j=40.5$ to $j=600$)
and some values
of the coupling parameter $\lambda$ ($0$, $0.25$, $0.5$, and $0.75$)
using standard matlab algorithms for sparse matrices.
For each spectrum we checked
that the mirror symmetry $E \mapsto -E$ is obeyed:
for $H_{\pm +}$
the mirror symmetry relates energies inside each subspectrum while
for $H_{\pm -}$ the mirror symmetry relates energies between the two subspectra.
If $E_\mathrm{max}>0$ is the largest 
and $E_\mathrm{min}<0$ the smallest eigenvalue
we estimate the mean level spacing by taking
\begin{equation}
  \overline{\Delta E}= \frac{E_{\mathrm{max}}-E_{\mathrm{min}}}{M-1}\ .
\end{equation}
In general larger values of $M$ lead to higher accuracy of the mean level 
spacing but also increase computation times, especially for large values of
$j$. Our choice is a compromise that we found reasonable.
The remaining analysis only uses a few positive energy eigenvalues
that we enumerate in increasing order $0< E_1 \le E_2 \le \dots$.
We unfold the spectrum by expressing each eigenvalue in units of the mean
level spacing
$e_n = \frac{E_n}{\overline{\Delta E}}$.
The average microscopic density of states is then defined by
\begin{equation}
  d(e)= \left\langle \sum_n \delta (e -e_n) \right\rangle
  \label{dos_definition}
\end{equation}
where $\langle \cdot \rangle$ is an appropriate averaging procedure.
In the standard Wigner-Dyson classes this procedure should lead to
$d(e) = 1$ up to numerical noise. In general the average may include
a disorder average. In the quantum chaos approach one often asks the question
whether an individual quantum system follows the random-matrix
prediction which is more challenging to show.
What this means in detail needs some clarification which we will give
below in Section~\ref{BGS} where we also explain that
an average over the total angular momentum  quantum number $j$
may be considered as an average for a single quantum system in a completely
analogous way to what is usually considered in quantum chaos.
Thus in order to avoid disorder we will consider averages of the form
\begin{equation}
  d(e) = \frac{1}{N_j}\sum_j \sum_{n} \delta_\epsilon(e-e^{(j)}_n) 
\label{average}
\end{equation}
where the sum is over $N_j$ different values of the total angular
momentum $j$ and corresponding energy spectra 
$\{ e^{(j)}_n \}_{n=1}^{n_{\mathrm{max}}}$ and only spectra that belong
to the same block $H_{\pm \pm}$ and the same symmetry class are added.
This is numerically a finite sum over Dirac-$delta$ peaks 
and we want to compare it to a smooth random-matrix prediction.
This may be achieved by using sufficiently broadened peaks 
with a finite width $\epsilon>0$
as indicated in \eqref{average} or, as we will do, by using 
appropriate integrated quantities.\\ 
Another quantity that shows clear quantum signatures of nonstandard
symmetries is the distribution of the smallest positive energy eigenvalue
\begin{equation}
  P_1(e)= \left\langle \delta(e-e_1) \right\rangle
  \equiv \frac{1}{N_j}\sum_{j} \delta_\epsilon(e-e_1^{(j)})\ .
\end{equation}

\subsection{The generalized Bohigas-Giannoni-Schmit conjecture }
\label{BGS}

As mentioned in the introduction Gaussian random-matrix ensembles
predict universal spectral statistics on the scale of mean level
spacing for quantum systems with a sufficient amount of complexity. 
In the case of disorder this is revealed by averaging over 
system parameters
and this approach is well established for all symmetry classes.
The Gaussian ensembles of random-matrix theory are indeed extreme disorder 
models where
all matrix elements that are not related by the underlying symmetries
are independent random variables with a Gaussian joint
probability density function $\propto e^{- A \mathrm{tr}\ H^2}$.

In the case of Quantum Chaos
the challenge is usually to avoid any average over disorder
and consider an \emph{individual} system.
This is included in the statement of the BGS conjecture 
(for standard Wigner-Dyson symmetry classes) that
quantum chaos (in the sense of full chaos in the corresponding classical
dynamics) implies universal spectral fluctuations as predicted
by the Gaussian random-matrix ensemble of the appropriate symmetry
class.
Before generalizing this statement to nonstandard symmetry classes
the statement of the BGS conjecture deserves clarification
especially
on what is meant by an \emph{individual} system.
In the standard symmetry classes the general idea is to replace
the disorder average by a semiclassical analysis of a spectral
average. Employing such a spectral average for an \emph{individual}
system may be misunderstood to imply that one should use
only a single quantum spectrum. We will argue that, in general, this
contradicts the semiclassical limit. 
In systems with a finite Hilbert space this is obvious as one
certainly needs an infinite spectrum to recover the smooth
predictions of random-matrix theory. If the system has
an infinite (discrete) spectrum one may divide it into subspectra
and create an ensemble. However the subspectra then belong to
different energy intervals. Unless the classical dynamics remains
unchanged up to trivial changes of scale these subspectra describe
different classical dynamics. Some paradigmatic models
of quantum chaos 
such as quantum billiards or particles in a scale-invariant potential
(see e.g.~\cite{berryrobnik}) are scale-invariant.
For these the semiclassical limit $\hbar \to 0$ (at fixed energy)
is equivalent to $E\to \infty$ (for a fixed value of $\hbar$)
and one may use a single infinite spectrum (divided into subspectra of
increasing size as $E$ increases) to generate an appropriate
ensemble that corresponds to a spectral average without disorder.\\
In more general systems that are either finite dimensional or not
scale invariant the semiclassical limit
$\hbar \to 0$ at fixed energy cannot be analyzed using a single
spectrum.
In these cases 
the semiclassical limit implicitly implies that one considers a 
sequence of spectra: as the formal value of $\hbar$ decreases the 
spectrum indeed changes. Changing the formal value of $\hbar$
physically means that other system parameters (such as masses,
coupling constants or field strengths) are changed and  
units are appropriately rescaled.
As $\hbar \to 0$ the size of the spectrum increases
and one may take the spectral average of a larger and larger spectrum
which may be compared to the random-matrix prediction.
Thus in quantum chaos, when one uses many quantum spectra 
for different values of $\hbar$ for an otherwise identical system 
one still refers to this as one \emph{individual} quantum system. 
While the reference to individual systems can be misleading
without clarification
there is some justification because the 
corresponding classical 
dynamics truly remains the same. Moreover such a sequence
defines a clear procedure
what is meant by analysing spectral statistics without using disorder.

What is seldom part of the standard procedure is to add
a running average over the formal value of $\hbar$ 
despite the fact that it would usually reduce noise in the limit.
The main reason for not using it  is probably that it is difficult
to implement the running average in any useful way in analytical
calculations. A second reason may be to obtain apparently stronger
results by avoiding any additional averaging process.
In the context of the BGS conjecture
for the standard symmetry classes it may be a matter of taste whether
one adds an average over $\hbar$. For nonstandard 
symmetry classes a running average over $\hbar$ 
seems to be the only way in which the spirit of the BGS
conjecture may be generalized. Any spectral average would wash out the
specific signatures of spectral fluctuations near $E=0$ that are related
to the nonstandard symmetry classes.
We thus suggest that the BGS conjecture is generalized
to the nonstandard symmetry classes by using the running
average over $\hbar$ with a fixed corresponding classical dynamics. 
This is consistent
with a previous generalized BGS conjecture for quantum graphs
with nonstandard symmetries \cite{stargraphI,stargraphII}.

\subsection{Numerical analysis of quantum signatures
related to nonstandard symmetries}

Let us now summarize the known random-matrix predictions
for the Gaussian ensembles (in the limit 
$N \to \infty$ of large matrix dimensions) 
for the relevant symmetry classes
present in the Feingold-Peres model and compare them to numerical
data.
For the orthogonal class A$I$,
where the ensemble is the well-known 
Gaussian orthogonal ensemble (GOE), 
the average microscopic density is just
\begin{equation}
  d(e)=1
\end{equation}
as in all Wigner-Dyson classes.
For the nonstandard symmetry classes the relevant
results may be found in  \cite{VS, Ivanov,edelmann}.
For the chiral orthogonal classes BD$I_\nu$
(with topological index $\nu$)  
the ensemble is known as the chiral 
Gaussian orthogonal ensemble or CHGOE$_{\nu}$
which gives
\begin{equation}
  d_{\mathrm{BD}I}(e)=
  \nu\delta(e)+
  \frac{\pi}{2}
  \left[
    \pi |e|
    \left(J_\nu(\pi |e|)^2-J_{\nu-1}(\pi|e|)J_{\nu+1}(\pi|e|)
    \right)
    +J_{\nu}(\pi|e|)\left(1-\int_0^{\pi|e|} J_\nu(z)dz \right)
  \right]\ .
\end{equation}
In the `antichiral' orthogonal class C$I$ 
the corresponding Gaussian ensemble does not have an established
name. It gives 
\begin{equation}
  d_{\mathrm{C}I}(e)=\frac{\pi}{2}\int_0^{\pi |e|}\frac{
    J_0(z)J_1(z)}{z} dz
  =\frac{\pi^2|e|}{2}\left(J_0(\pi |e|)^2 + J_1(\pi |e|)^2 \right)-
  \frac{\pi}{2}J_0(\pi |e|)J_1(\pi |e|)
\end{equation}
and, apart from the $\delta$ peak at $e=0$,  
this result coincides with the prediction 
for class BD$I_1$.\\
All predictions satisfy
\begin{equation}
  d_{\mathrm{BD}I_\nu}(e)\rightarrow 1 \quad \text{and} \quad
  d_{\mathrm{C}I}(e)\rightarrow 1 \quad \text{as} \quad |e| \to \infty\ .
\end{equation}
Close to $e=0$ the average density 
of states is increased for BD$I_0$ (eigenvalues are attracted to $e=0$)
while for BD$I_1$ and C$I$ the density of states is suppressed 
with $d(e) \to 0$ as $e\to 0$ (eigenvalues repelled from $e=0$).
Let us write
\begin{equation}
  d(e)=1+\nu \delta(e) + \delta d(e)
\end{equation}
\begin{figure}
  \includegraphics[width=0.4\textwidth]{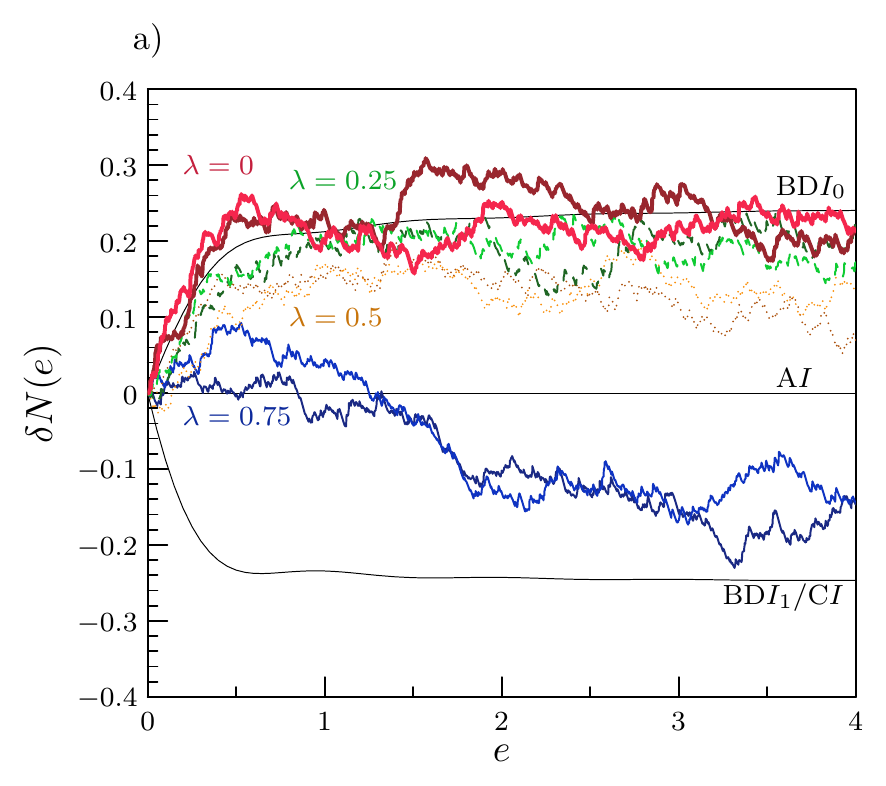}
  \includegraphics[width=0.4\textwidth]{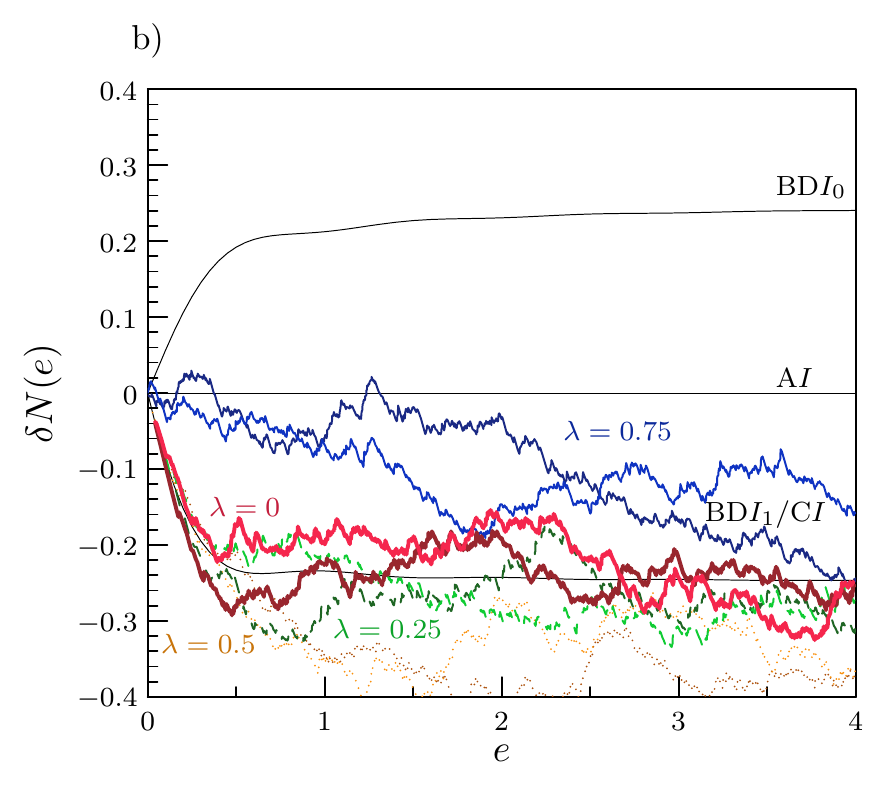}\\
  \includegraphics[width=0.4\textwidth]{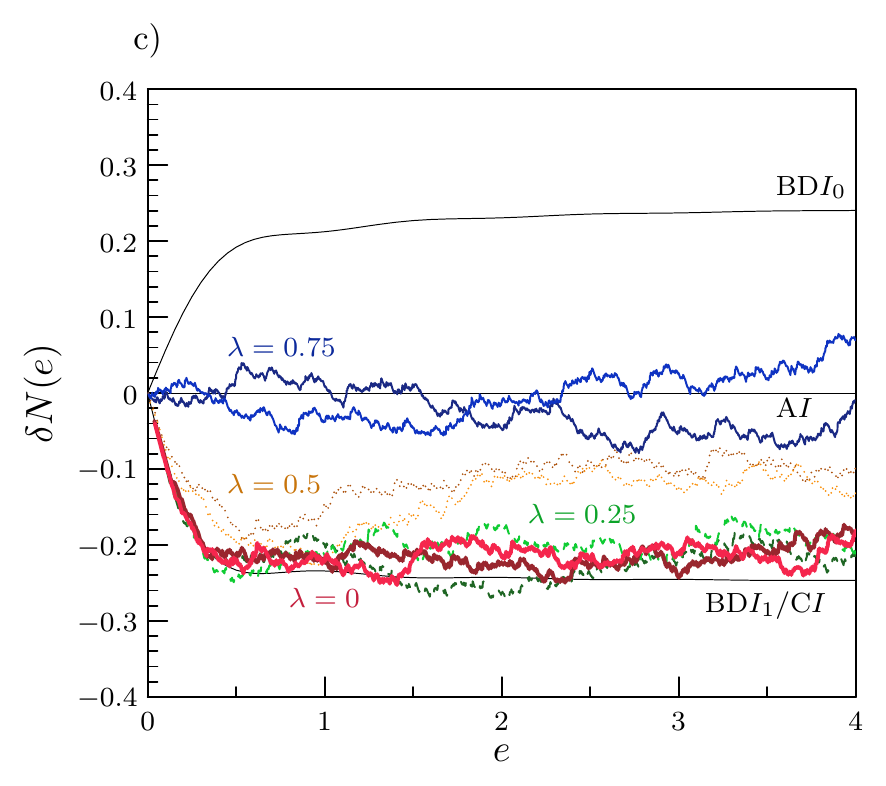}
  \includegraphics[width=0.4\textwidth]{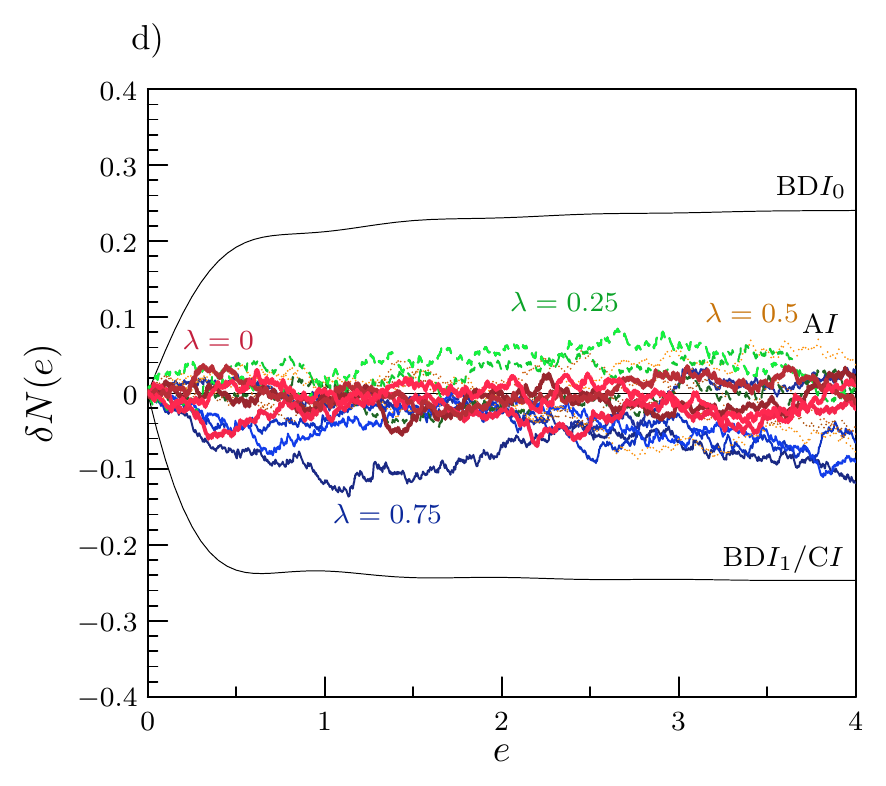}
  \caption{(Color online) 
    Integrated deviation of the microscopic density of states 
    for different values of the coupling parameter
    (fat red lines for $\lambda=0$, dashed green lines for $\lambda=0.25$,
    dotted yellow lines for $\lambda=0.5$ and thin 
    blue lines for $\lambda=0.75$).
    Thin black lines correspond to random-matrix predictions for the
    indicated symmetry classes.
    Graph a) shows the spectra of the blocks $H_{++}$ and $H_{-+}$ 
    that belong to symmetry class BD$I_0$, i.e. for integer $j$ and even
    dimensional invariant subspaces (each curve combines 280 spectra). 
    Graph b) shows spectra of the same blocks  $H_{++}$ and $H_{-+}$ that 
    belong to symmetry class 
    BD$I_1$, i.e. integer $j$ and odd dimensional subspaces 
    (again 280 spectra per line). Graph c) shows the
    spectra of $H_{++}$ and $H_{-+}$ in symmetry class C$I$, i.e.
    half-integer $j$ (560 spectra per line). Graph d) shows the spectra
    for the blocks $H_{+-}$ and $H_{--}$ that belong to the standard
    Wigner-Dyson class A$I$ where integer and half-integer values of $j$ have
    been combined separately
    (560 spectra per line).
    \label{fig3}}
\end{figure}
such that all nontrivial signatures are contained in $\delta d(e)$.
In Figure~\ref{fig3} we plot the integrated version 
\begin{equation}
  \delta N(e) =\int_0^e \delta d(e') \ de'.
\end{equation}
The graphs support the generalized BGS conjecture in that they
follow the symmetry class specific signatures as predicted by
the Gaussian random-matrix ensemble for the corresponding
symmetry classes if the classical dynamics is chaotic
at $\lambda =0 $ and $\lambda=0.25$.
The signatures become weaker if the classical dynamics is mixed
$\lambda=0.5$ and disappear or change further in the near-integrable case
$\lambda=0.75$. 
We do not show 
the fully integrable case $\lambda=1$ which is highly 
degenerate and ungeneric both classically and quantum mechanically.
For the 
spectra in the standard symmetry classes no signatures are expected
irrespective of the classical dynamics. This is confirmed in 
Figure~\ref{fig3} d).

Let us next turn to the distribution $P_1(e)$ of the first positive 
eigenvalue.
Numerically we will consider the integrated distribution
\begin{equation}
  I(e)=\int_0^e P_1(e') de'.
\end{equation}
$E(e)=1-I(e)$ is also known as the gap probability.
In the three Wigner-Dyson classes this can be calculated directly from 
the level spacing distribution $P(s)$ 
\begin{equation}
  I(e)=\int_0^e s P(s) ds + e\int_e^\infty P(s) ds
\end{equation}
For the
GOE the level-spacing distribution is given by
Wigner surmise (up to very small corrections
that are beyond our numerical analysis)
\begin{equation}
  P(s)=\frac{\pi}{2} \exp \left({-\frac{\pi}{4} s^2}\right) \ .
\end{equation}
This leads to
\begin{equation}
  I_{\mathrm{GOE}}(e)= \mathrm{erf} \left(\frac{\sqrt{\pi}}{2} e\right)
\end{equation}
which is the prediction for quantum chaotic spectra
in class A$I$ (modulo above mentioned tiny corrections). Integrable
systems in any of the  Wigner-Dyson classes have Poissonian spectra
with $P(s)=e^{-s}$. The prediction for integrable
systems in class A$I$ is then
\begin{equation}
  I_{\mathrm{Poisson}}(e)= 1-\exp(-e)\ .
\end{equation}
For $e\to 0$ both, the GOE and the Poisson prediction behave like
$I(e) \sim e$ which is consistent with a featureless density
of states $d(e)=1$.
The stronger spectral rigidity of quantum chaotic spectra 
versus integrable systems can be seen from the much quicker saturation
$I_{\mathrm{GOE}}(e) \to 1$ compared to the Poissonian case.
In the nonstandard symmetry classes predictions for quantum chaotic
spectra are obtained from the 
well-known joint-probability distributions of eigenvalues
of the corresponding Gaussian ensembles of
random-matrix theory (a 
very useful overview of all relevant information can be found in the 
Appendix of \cite{kieburg}). 
In class BD$I_0$ one has
\begin{equation}
  I_{\mathrm{BD}I_0}(e)=1-\exp\left(-\frac{\pi^2}{8} e^2- \frac{\pi}{2} e\right)\ .
\end{equation}
The attraction of eigenvalues to $e=0$ shows in the steeper
behavior $I_{\mathrm{BD}I_0}(e) \sim \frac{\pi}{2} e$.
In the classes BD$I_1$ and C$I$ one has (again) the same prediction
\begin{equation}
  I_{\mathrm{BD}I_1}(e)=  I_{\mathrm{C}I}(e)=  1-\exp\left(-\frac{\pi^2}{8}e^2\right)
\end{equation}
which show level repulsion from $e=0$ as 
$I(e) \sim 
\frac{\pi^2}{8}e^2$ only grows quadratically.
\begin{figure}
  \includegraphics[width=0.4\textwidth]{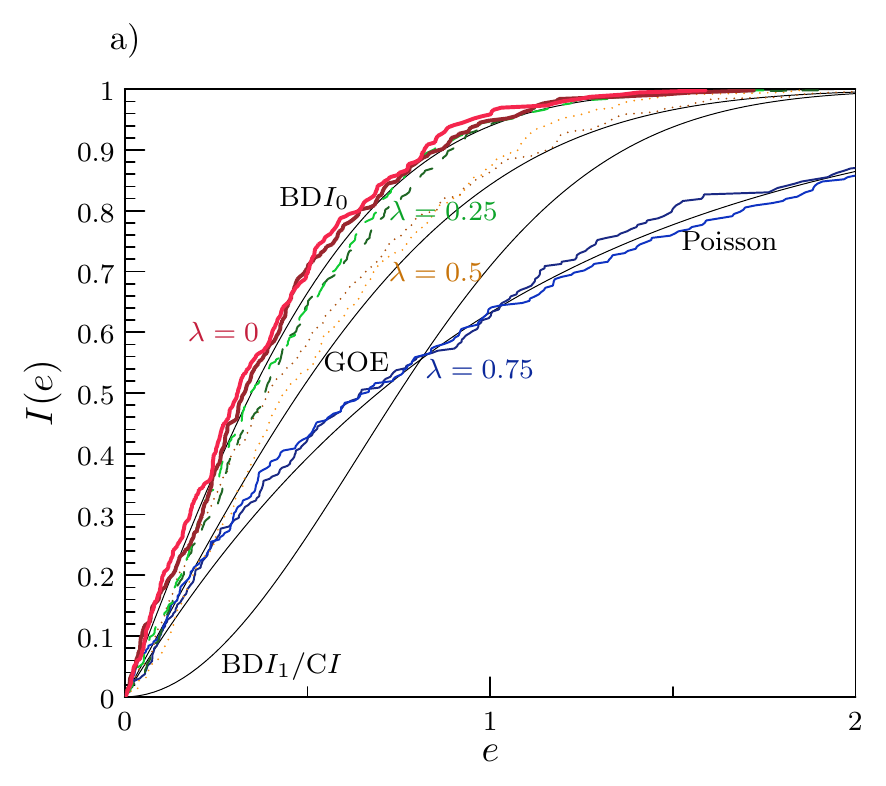}
  \includegraphics[width=0.4\textwidth]{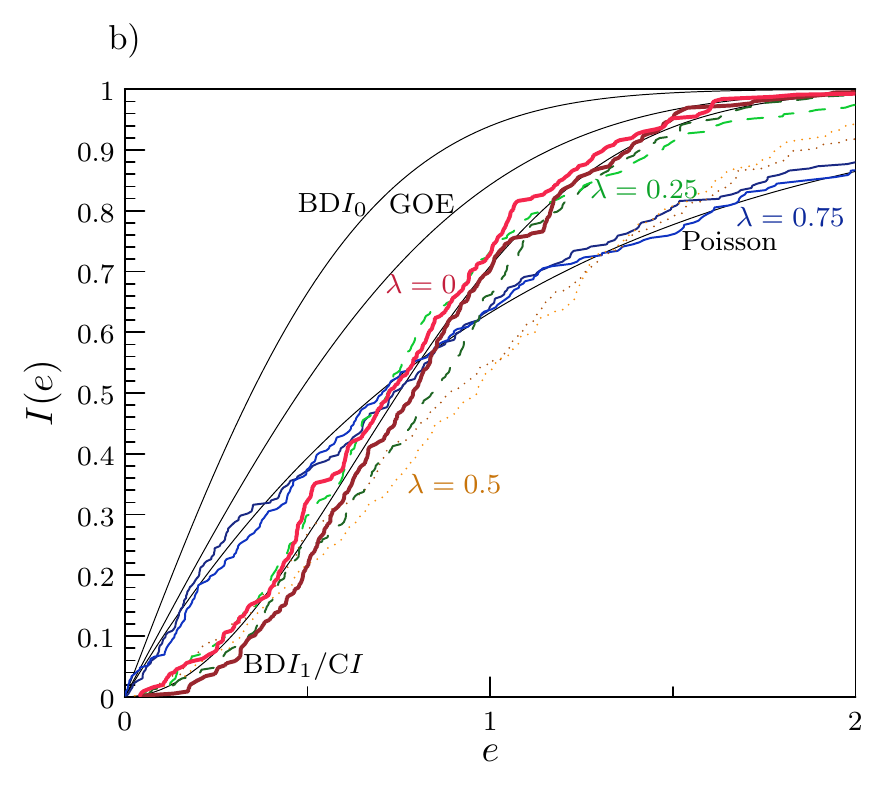}\\
  \includegraphics[width=0.4\textwidth]{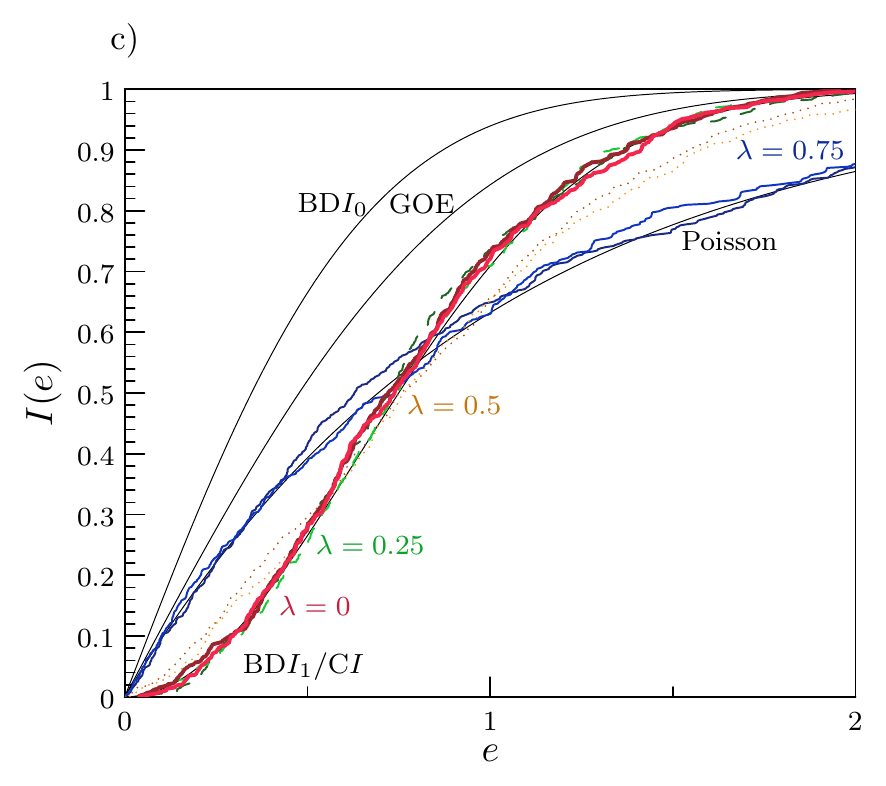}
  \includegraphics[width=0.4\textwidth]{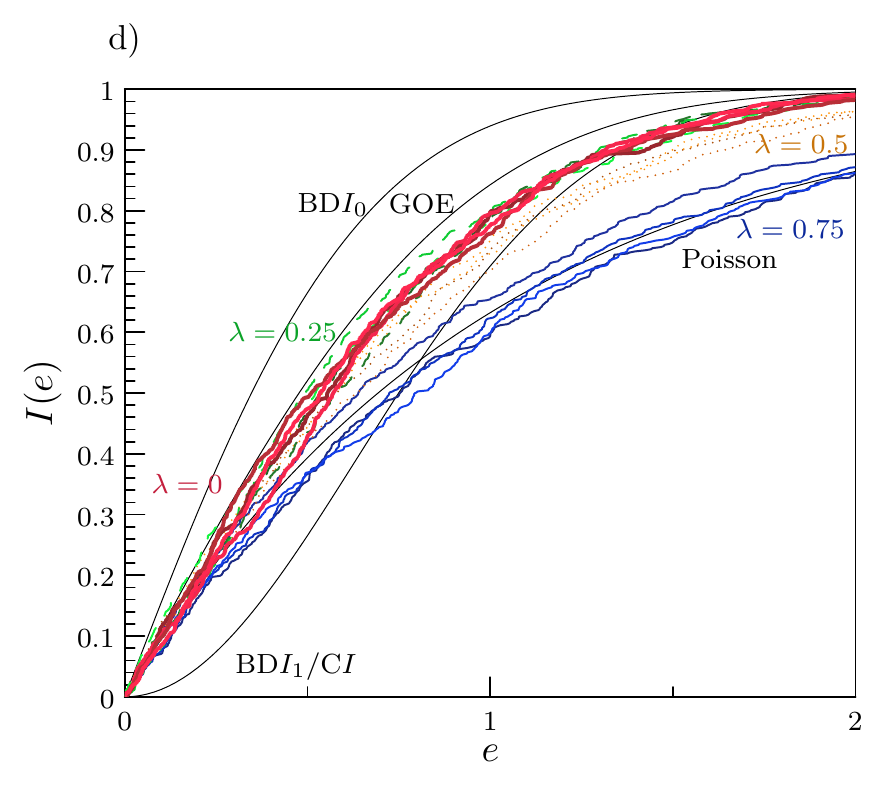}
  \caption{(Color online) Integrated distribution of the first positive eigenvalue.
    The description of individual curves and graphs 
    of Fig.~\ref{fig3} applies here analogously. 
    \label{fig4}}
\end{figure}
In Figure~\ref{fig4} we compare the random-matrix predictions
to the numerical data for the various symmetry classes realized in different
subspaces. Again we find support of the generalized BGS conjecture.
The spectra in quantum chaotic cases $\lambda=0$ and $\lambda=0.25$ 
follow the corresponding random-matrix prediction and we see a crossover
to Poissonian statistics in the mixed dynamics case $\lambda=0.5$ 
and the near-integrable case $\lambda=0.75$.
The distinct signatures of the nonstandard classes are very clear
as the fluctuations are much smaller than the difference between
the various predictions.

\section{Conclusions}
\label{sec:conclusions}

We have shown that the Feingold-Peres model of two coupled tops
belongs to nonstandard symmetry classes and we numerically
analyzed their spectra in a crossover where the corresponding classical
dynamics changes from chaotic to mixed and near-integrable.
For chaotic classical dynamics the spectral fluctuation near 
the spectral symmetry point $E=0$ 
show the characteristic signatures  
of nonstandard symmetry classes as predicted by the corresponding
Gaussian ensembles of random-matrix theory:
levels are repelled or attracted to $E=0$ depending
on the nonstandard symmetry class while
the standard Wigner-Dyson classes do not have a spectral symmetry
point and show no attraction or repulsion from $E=0$.
The signatures can very clearly be seen by looking
at the density of states (or its integral the spectral counting
function) or the (integrated) distribution of the first positive eigenvalue.
For nonchaotic classical dynamics these signatures either disappear or
change drastically.\\
This gives numerical
support for a generalized BGS conjecture, that chaos in
the classical dynamics implies universal spectral statistics
that only depends on the symmetry classification according to the 
Altland-Zirnbauer tenfold way. 
Our analysis is a further step in understanding
the specific spectral signatures related to nonstandard symmetry classes.
Two obvious directions for future work are
\begin{enumerate}
\item the extension to other symmetry classes that
  are not covered by the Feingold-Peres Hamiltonian \eqref{Hamiltonian},
\item the semiclassical analysis based on 
  periodic-orbit theory in the Gutzwiller trace formula. 
\end{enumerate}
Let us mention a few words on both directions.\\
Currently there is only a restricted set of paradigm models
for quantum chaos in nonstandard symmetry classes available in the literature.
Magnetic Andreev billiards provide a model in class C and the present work
adds models for classes BD$I$ and C$I$. It is in principle not difficult
to construct Hamiltonians for two couples tops that generalize
the Feingold-Peres model to other nonstandard symmetry classes.
For such a constructions one may make a choice of the appropriate chirality,
charge conjugation or time-reversal operators for a given symmetry class.
Then one writes down all monomial combinations of the angular momentum
components that are consistent with the symmetry and chooses a linear
combination as a family of Hamiltonians.
One may also generalize by allowing the total angular momentum quantum numbers 
of the two tops to be different $\mathbf{L}^2\equiv j_1(j_1+1)$
and $\mathbf{M}^2 \equiv j_2 (j_2 +1)$. As an example let us 
chose the standard time reversal operator, i.e. the
antiunitary operator $T$ that reverses each angular momentum
$T \mathbf{L} T^{-1}= - \mathbf{L}$ and $T \mathbf{M} T^{-1}= - \mathbf{M}$.
Then $T^2=(-1)^{2(j_1+j_2)}= \pm 1$ depending on whether $j_1$ and $j_2$
are both integer ($T^2=1$) both half-integer (also $T^2=1$)
or one is integer and the other half-integer ($T^2=-1$).
Next let us choose the unitary chirality operator 
$C= e^{i\pi M_x}$ which satisfies $TCT^{-1}=C$ and obeys
$C^2=(-1)^{2j_2}= \pm 1$ depending on whether 
$j_2$ is integer ($C^2=1$) or half-integer ($C^2=-1$).
The set of Hamiltonians
\begin{equation}
  H= \frac{M_y}{j_2+1/2}\frac{a_x L_x+ a_y L_y + a_z L_z}{j_1+1/2} 
  +\frac{M_z}{j_2+1/2}\frac{b_x L_x+ b_y L_y + b_z L_z}{j_1+1/2}
  + \frac{M_x\left(c_yM_y + c_z M_z \right)}{(j_2+1/2)^2}
\end{equation}
where $a_x$, \dots, $b_x$, \dots, $c_y$ and $c_z$ are real coupling
constants
then obeys $CHC^{-1}= -H$ and $THT^{-1} =H$.
Any unitary symmetry may be broken by changing some of the
coupling constants such that further reduction is generally 
not necessary.
Four different nonsymmetry classes are realized generically
depending on $j_1$
and $j_2$: if $j_1$ and $j_2$ are both integer the symmetry class is  
again BD$I$, if they are both half-integer then it is C$I$,
if $j_1$ is integer but $j_2$ half-integer, then we have D$III$, and if
$j_1$ is half-integer and $j_2$ integer, then one gets C$II$.
It is not difficult to see that all ten symmetry classes can
be realized by generalizing the Feingold-Peres model in this way.
This leads to a large set of potential paradigm models.
The corresponding classical limit is obtained by taking $j_1 \to \infty$
and $j_2 \to \infty$ in an appropriate way. However, 
any detailed future quantum chaos analysis
of these models may have to start with scanning the parameter space
for full classical chaos or mixed dynamics.\\
The second direction mentioned above is the semiclassical
periodic orbit approach based on the Gutzwiller
trace formula which expresses the microscopic density of states
as a sum
\begin{equation}
  d(e)= 1 + \left\langle
    \sum_p A_p e^{i S_p/\hbar} \right \rangle
\end{equation}
of a Weyl part that reduces to the unit constant 
(energies are measured in units of the mean level spacing) and
a sum over periodic orbits $p$ of the corresponding classical dynamics.
Each periodic orbit contributes a complex amplitude $A_p e^{i S_p/\hbar}$
where $S_p$ is the classical action $S_p= \oint \mathbf{P} \cdot d \mathbf{Q}$
of the periodic orbit and $A_p$ may be expressed in terms of 
the classical stability of the orbit. The average $\langle \cdot \rangle$
is a running average over values of $\hbar$ while one consider the 
semiclassical limit $\hbar \to 0$. Only periodic orbits whose total action
is very small can contribute. 
For standard symmetry classes there is no mechanism for getting
arbitrarily small classical actions and the sum over periodic orbit
does not survive the average. In the presence of a nonstandard symmetry
there is however a classical transformation corresponding to the quantum
chiral or charge conjugation symmetry. This maps
the energy shell $e=0$ into itself such each periodic orbit $p$
in the shell $e=0$ is mapped into a partner orbit $p'$. If $p=p'$
the periodic orbit is called \emph{self-dual}. For magnetic Andreev billiards
\cite{MagneticAndreev}
it has been shown that the total action $S_p$ of a self-dual
orbit vanishes and that semiclassical sum rules may be applied to sum
over all amplitudes $A_p$ of self-dual orbits.
The resulting \emph{self-dual approximation} is analogous to
Berry's well-known diagonal approximation for the 
semiclassical two-point correlation 
function \cite{diagonal}. The self-dual approximation can be performed in all
nonstandard symmetry classes for quantum graphs \cite{stargraphII}.
In the present case of two coupled tops 
it is not immediately obvious that the self-dual approximation
is consistent with the Gaussian random-matrix predictions. We hope that
future work will clarify this in detail.

\appendix
\section{Explicit derivation of the topological index}
\label{appendix}

In this appendix we assume that $j$ is integer and show that the
topological 
index of the reduced Hamiltonian $H_{++}$ is $\nu_{++}=0$ if $j$ is
odd and $\nu_{++}=1$ if $j$ is even. The corresponding statement for 
$H_{-+}$ ($\nu_{-+}=0$ if $j$ is
even and $\nu_{-+}=1$ if $j$ is odd) may be proven in analogously.\\
From $C|m_1, m_2\rangle=(-1)^{j-m_2}|-m_1,-m_2\rangle$ 
we can find the
action
of the reduced chirality operator $C_{++}$ as
\begin{equation}
  C_{++}|m_1,m_2,++\rangle= (-1)^{j-m_2}|-m_2,-m_1,++\rangle
  \label{cpp_action}
\end{equation} 
where we have used that $m_1+m_2$ is even in the subspace
$V_{++}$. If $m_1\neq -m_2$, then we find the eigenstates
\begin{subequations}
  \begin{align}
    C_{++}\left( |m_1,m_2,++\rangle + |-m_2,-m_1,++\rangle\right)=&
    =(-1)^{j-m_2}\left( |m_1,m_2,++\rangle + |-m_2,-m_1,++\rangle\right)\\
    C_{++}\left( |m_1,m_2,++\rangle - |-m_2,-m_1,++\rangle\right)=&
    =(-1)^{j-m_2+1}\left( |m_1,m_2,++\rangle - |-m_2,-m_1,++\rangle\right)
  \end{align}
\end{subequations}
which have pairwise eigenvalues $\pm 1$ and so do not contribute to
the topological index.\\
If $m_1=-m_2$ then \eqref{cpp_action} implies 
\begin{equation}
  C_{++}|m_1,-m_1,++\rangle= (-1)^{j+m_1}|m_1,-m_1,++\rangle
\end{equation}
with eigenvalues $(-1)^{j+m_1}$. As $m_2\ge m_1$ and $m_1+m_2$ are
even for $V_{++}$ where $m_1$ runs over the $j+1$ values
$m_1=-j,-j+1,\dots,0$. The alternating signs of the eigenvalues then
imply
that the topological index is $\nu_{++}=1$ if $j$ is even, and 
$\nu_{++}=0$ if $j$ is even as claimed.

\begin{acknowledgments}
  We thank Gernot Akemann for helpful
  discussion and pointing us to helpful literature.
  Yiyun Fan and Yuqi Liang have been supported 
  through a Summer Research Bursary by the School of Mathematical
  Sciences, University of Nottingham and a Dr
  Margaret Jackson Bursary Award. 
\end{acknowledgments}


\begin{thebibliography}{10}

\bibitem{fp} M.~Feingold, A.~Peres, Physica \textbf{9D}, 433 (1983).
\bibitem{fmp} M.~Feingold, N.~Moiseyev, A.~Peres, Phys.~Rev.~A \textbf{30}, 509 (1984).
\bibitem{AZ1} A.~Altland, M.R.~Zirnbauer, Phys.~Rev.~Lett. \textbf{76}, 
3420 (1996).
\bibitem{AZ2}
  A.~Altland, M.R.~Zirnbauer, Phys.~Rev.~B \textbf{55}, 1142 (1997).
\bibitem{Zirni} M.R.~Zirnbauer, J.~Math.~Phys.~\textbf{37}, 4986 (1996).
\bibitem{Wigner1} E.P.~Wigner, Proc.~Cambridge~Philos.~Soc. \textbf{47}, 790 (1951)
\bibitem{Wigner2}
E.P.~Wigner, Ann.~Math.~\textbf{67}, 325 (1958).
\bibitem{Dyson1}
F.J.~Dyson, J.~Math.~Phys.~\textbf{3}, 140 (1962).
\bibitem{Dyson2}
F.J.~Dyson, J.~Math.~Phys.~\textbf{3}, 1199 (1962).
\bibitem{Ver1}
  J.J.M. Verbaarschot, I. Zahed, Phys.~Rev.~Lett. \textbf{70}, 3852
  (1993).
\bibitem{Ver2}
  J.J.M. Verbaarschot, Phys.~Rev.~Lett. \textbf{72}, 2531 (1994).
\bibitem{gade}
  R.~Gade, Nucl.~Phys.~B~\textbf{398}, 499 (1993).
\bibitem{slevin} K.~Slevin, T.~Nagao,
  Phys.~Rev.~Lett. \textbf{70}, 635 (1993).
\bibitem{wishart} J.~Wishart, Biometrika \textbf{20A}, 32 (1928).
\bibitem{stargraphI} S.~Gnutzmann, B.~Seif, Phys.~Rev.~E~\textbf{69}, 
  056219 (2004).
\bibitem{BGS} O.~Bohigas, M.J.~Giannoni, C.~Schmit, 
  Phys.~Rev.~Lett. \textbf{52}, 1 (1984).
\bibitem{diagonal} M.V.~Berry, Proc.~R.~Soc.~Lond.~A
  {\bf 400}, 299 (1985).
\bibitem{actioncorrelation} N.~Argaman, F.M.~Dittes, E.~Doron, J.~Keating, 
  A.~Kitaev, M.~Sieber, U.~Smilansky, 
  Phys.~Rev.~Lett. {\bf 71}, 4326 (1993).
\bibitem{SieberRichter}  M.~Sieber, K.~Richter, Phys. Scr. {\bf T90},
  128 (2001).
\bibitem{GA1}
  S.~Gnutzmann, A.~Altland, Phys.~Rev.~Lett. \textbf{93}, 194101 (2004).
\bibitem{GA2}
  S.~Gnutzmann, A.~Altland, 
  Phys.~Rev.~E~\textbf{72}, 056215 (2005).
\bibitem{quantumgraphs} 
  S.~Gnutzmann, U.~Smilansky, Adv. Phys. \textbf{55}, 527 (2006).
\bibitem{Essen}
  S.~M\"uller, S.~Heusler, A.~Altland, P.~Braun, F.~Haake,
  New J.~of Phys.~\textbf{11}, 103025 (2009).
\bibitem{SuSyreview} 
  A.~Altland, S.~Gnutzmann, F.~Haake, T.~Micklitz,
  Rep.~Prog.~Phys.~\textbf{78}, 086001 (2015).
\bibitem{catmaps}
  J.P. Keating,  Nonlinearity~\textbf{4}, 309 (1991).
\bibitem{arithmetic}
  E.~B. Bogomolny, B.~Georgeot, M.~J. Giannoni, C.~Schmit,
  Physics Reports \textbf{291}, 219 (1997).
\bibitem{berryrobnik} M.V. Berry, M.~Robnik M,  J. Phys. A
\textbf{17}, 2413 (1984).
\bibitem{stargraphII} S.~Gnutzmann, B.~Seif, Phys.~Rev.~E~\textbf{69}, 
  056220 (2004).
\bibitem{MagneticAndreev} S.~Gnutzmann, B.~Seif, F.~von Oppen, 
  M.R.~Zirnbauer, Phys.~Rev.~E\textbf{67}, 046225 (2003).
\bibitem{kosztin}
  I.~Kosztin, D.L.~Maslov, P.M.~Goldbart,
  Phys.~Rev.~Lett. \textbf{75}, 1735 (1995).
\bibitem{AndreevReview}
  C.W.J. Beenakker, Lecture Notes in Physics \textbf{667}, 131-174
  (2005).
\bibitem{VS}
J.J.M.~Verbaarschot, Nucl.Phys. \textbf{B426}, 559  (1994).
\bibitem{Ivanov}
  D.A.~Ivanov, J.~Math.~Phys. \textbf{43}, 126 (2002).
\bibitem{edelmann}  A.~Edelman,
  Lin.~Alg.~Appl. \textbf{159}, 55 (1991).
\bibitem{kieburg} M.~Kieburg, T.R.~W\"urfel, Phys.~Rev.~D \textbf{96},
  034502 (2017).
\end{thebibliography}
\end{document}